\documentclass [12pt]{article}   
\usepackage{array}
\title{Normalization Constants of Large Order Behavior}

\author{Taekoon Lee\footnote{email: tlee@ctp.snu.ac.kr}
        \\
        \\
        Center for Theoretical Physics\\
        Seoul National  University\\
        Seoul 151-742,  Korea}

\date{}
\begin{document}
\maketitle
\begin{abstract}
A perturbation scheme is discussed for the computation of the
normalization constant of the large order behavior arising from an ultraviolet
renormalon. In this scheme the normalization constant is expressed in
a convergent series that can be calculated using the ordinary 
perturbative expansion and the nature of the renormalon singularity. 
\end{abstract}

\vspace{2.5in}
\begin{tabbing}
{\it PACS numbers: 11.15.Bt, 11.10.Jj, 11.25.Db} \\
{\it Keywords: renormalon, large order behavior}
\end{tabbing}

\def\thepage{SNUTP-99-039}
\thispagestyle{myheadings}
\newpage
\pagenumbering{arabic}
\addtocounter{page}{0}
\newcommand{\be}{\begin{equation}}
\newcommand{\ee}{\end{equation}}
\newcommand{\bear}{\begin{eqnarray}}
\newcommand{\eear}{\end{eqnarray}}

The large order behavior in field theories arising from a renormalon
is generally
given in the form
\begin{equation}
a_{n}=K n! n^{\nu} b_{0}^{-n}[ 1+ O(1/n)]  
\hspace{.25in} \mbox{for}\hspace{.25in} n \rightarrow \infty.
\label{eq1}
\end{equation}
While the constants $\nu$ and $b_{0}$ are calculable, the normalization
constant $K$ cannot be determined exactly. An infinite number of
renormalon diagrams contribute to it, but it is not known how to sum
such diagrams to all order \cite{tlee,grunberg,zakharov}.

Though the normalization constant cannot be determined exactly, we shall see
that it can be calculated perturbatively to an arbitrary precision.
The large flavor ($N_{f}$) expansion is often invoked  for an
approximate evaluation of the normalization, and in the literature  it is
further asserted  that the normalization cannot be computed without
resorting to it \cite{beneke}. However,  the $N_{f}$ expansion
may not be considered a systematic perturbation scheme,
as it is not proven that it gives a convergent series or even
compatible with nonabelian gauge theories. At large $N_{f}$, the asymptotic
freedom is lost, and so it is probably incompatible with asymptotic free
gauge theories.

We note, however, that there is a systematic method that is both compatible
with nonabelian gauge theories and gives a convergent series.
In \cite{tlee} we have discussed a  scheme  which express
the normalization constant of an infrared (IR) renormalon in a convergent 
series which depends only  on the strength of the renormalon
singularity and the ordinary perturbative coefficients of the amplitude in
consideration. A sample calculation in QCD using the radiative calculations
up to three loops shows that our method gives a rather quickly convergent
series. 
The purpose of this letter is to extend the scheme to the case of
an UV renormalon.

Let us  first  review the perturbation scheme briefly in the case of an
IR renormalon. To be specific, we  consider a Green's function $D(\alpha)$,
such as the Adler function, in QCD and its expansion
\begin{equation}
D(\alpha)= \sum_{n=0}^{\infty} a_{n} \alpha^{n+1}.
\end{equation}
The Borel transform  $\tilde{D}$ is then defined as follows. We first define 
it in the neighborhood of the origin as
\begin{equation}
\tilde{D}(b)=\sum_{n=0}^{\infty} \frac{a_{n}}{n!} b^{n},
\end{equation}
and then by analytically continuing to the whole $b$-plane.
$\tilde{D}(b)$ is known to have IR renormalon singularities at
 $ b= -n/\beta_{0}$, $n= 2,3,4,\cdots$ and UV renormalon singularities at
$ b=  n/\beta_{0}$, $n= 1,2,3,\cdots$, where
$\beta_{0}$ is the first coefficient of the $\beta$-function. Throughout
this letter we assume there are no other singularities associated with
$\tilde{D}(b)$ except for those caused by the instantons which are irrelevant
for our discussion. Now consider the first IR renormalon at
$ b=  -2/\beta_{0}$.
The nature of the first IR renormalon singularity is given in the form
\cite{mueller}
\begin{equation}
\tilde{D}(b)\approx \frac{\hat{D}}{\left(1+\frac{\beta_{0}b}{2}\right)^{1+\nu}}
\hspace{.25in} \mbox{for}\hspace{.25in} b\approx \frac{-2}{\beta_{0}},
\end{equation}
where $\nu= -2 \beta_{1}/\beta_{0}^{2}$, and $\beta_{1}$ is the second
coefficient of the $\beta$-function.
Then the large order behavior caused by the renormalon is given in the form
(\ref{eq1}) with $ K= \hat{D}/\nu!$ and $b_{0}=-2/\beta_{0}$.
Note that the normalization constant
becomes the residue of the singularity in the $b$-plane. Thus 
we can equivalently  work with  the renormalon  residue in order to study the
normalization constant.

The perturbative calculation of the residue is based on the
observation that the residue of the closest singularity  to the origin
in the complex  plane  can be expressed in a convergent series involving
only the Taylor expansion at the origin. 
Assuming that the strength of the singularity is known all one needs 
to calculate the residue is to move the singularity  in consideration 
by conformal mapping closer to the origin than any other singularities in 
the Borel plane.

This scheme works as follows in the case of the first IR renormalon.
The closest singularity to the origin in the $b$-plane is the
UV renormalon at $b =1/\beta_{0}$. Using the conformal mapping
\begin{equation}
z=-\frac{ \beta_{0} b}{1-\beta_{0}b}
\end{equation}
we can make the first IR renormalon the closest singularity to the origin.
In the $z$-plane the first IR renormalon assumes
\begin{equation}
\tilde{D}(b(z)) \approx \frac{ \left(\frac{2}{9}\right)^{1+\nu}\hat{D}}{
\left(\frac{2}{3}-z\right)^{1+\nu}}\, \hspace{.25in} \mbox{for}\hspace{.25in}
z \approx \frac{2}{3}.
\end{equation}
Now consider a function defined by
\begin{equation}
R(z)=\tilde{D}(b(z)) \left(\frac{2}{3}-z\right)^{1+\nu}.
\end{equation}
At the IR renormalon singularity,  $R(z)$ could  still  be singular,
but it is bounded.
The residue is then given by
\begin{equation}
\left(\frac{2}{9}\right)^{1+\nu}\hat{D}=R(z)\left.\right|_{z=\frac{2}{3}}.
\label{eq8}
\end{equation}
Since  $R(z)$ is  analytic on the disk $|z|<2/3$  we can write
(\ref{eq8}) in a series  form by expanding it at $z=0$
\begin{equation}
\left(\frac{2}{9}\right)^{1+\nu}\hat{D}=\sum_{n=0}^{\infty}
r_{n}z^n\left.\right|_{z=\frac{2}{3}}.
\label{eq9}
\end{equation}
It is easy to see that $ r_{n}$ depends only on $a_{i}$ with $i \leq n$,
and thus calculable. 

One might question the convergence of the series (\ref{eq9}), since
it is evaluated at the renormalon singularity at which $R(z)$ could be
singular. However, it should be noted that the finiteness of $R(z)$
at the singularity guarantees the convergence.
A numerical evaluation of the series using the 3-loop calculation of
the Adler function in QCD shows a quick convergence for
small $N_{f}$ case \cite{tlee}. For example, the first three
elements of the series are 0.904, -0.358, 0.003 for $N_f=2$, and 
0.946, -0.354, -0.098 for $N_f=3$.

This example demonstrates that a renormalon residue can be expressed
in a convergent series once the nature of the singularity is
known. We now show that a similar conclusion can be made with a UV
renormalon. In the following we shall 
assume that $\beta_{0} < 0$, for definiteness, and focus exclusively on the
first UV renormalon since it gives the dominant large order behavior.
The structure of the UV renormalon is a little more complicated. According
to Parisi, it is determined by an insertion of dim=6 operators \cite{parisi}.
To be specific, let us consider a Green's function $A(\alpha)$ and
its Borel transform $\tilde{A}(b)$. Generally the Borel transform has a
branch cut beginning at the first UV renormalon on the negative real axis,
and a quantity defined by
\begin{equation}
\mbox{Im} A(\alpha) = -\lim _{\epsilon\rightarrow 0} \frac{1}{2i}
\int_{-\infty}^{0} d\, b\, e^{-\frac{b}{\alpha}} \left[
\tilde{A}(b +i \epsilon) -\tilde{A}(b -i \epsilon) \right]
\end{equation}
is nonvanishing.
The Parisi ansatz states that the dominant contribution to
$\mbox{Im} A(\alpha)$,
for $\alpha \rightarrow 0_{-}$, arises from an insertion of dim=6 operators,
and is given in the form,
\begin{equation}
\mbox{Im} A(\alpha) = \sum_{i=1}^{M} f_{i}(\alpha) O_{i}(\alpha) +
O( e^{-\frac{2}{\beta_{0} \alpha}})
\end{equation}
where the index $i$ runs over all  dim=6 operators $O_i$.
And $f_{i}$ satisfy the
renormalization group equation
\begin{equation}
\left[ (\beta(\alpha) \frac{d}{d\alpha} -1)\delta_{ij}
-\gamma_{ij}(\alpha)\right] f_{j}(\alpha) =0
\end{equation}
where $\gamma_{ij}$ denotes the anomalous dimension of the associated dim=6
operators. An explicit implementation of the Parisi ansatz  may be 
found in \cite{braun}.
Solving the RG equation one can  write formally
\begin{equation}
\mbox{Im} A(\alpha) = -\sum_{i}^{M} \frac{\pi\beta_{0} K_{i}}{\Gamma(\nu_{i})
}e^{-\frac{1}{\beta_{0} \alpha}}
\left(\frac{\alpha}{\beta_{0}}\right)^{1-\nu_{i}} \left[1 +
\sum_{j=1}^{\infty}\frac{\Gamma(\nu_{i})C_{ij}}{\Gamma(\nu_{i}-j)}
\left(\frac{\alpha}{\beta_{0}}\right)^{j}\right] +O( e^{-\frac{2}{\beta_{0}
\alpha}})
\label{eq13}
\end{equation}
where $K_{i}$ are undetermined constants while $\nu_{i}$ depend on both
$\beta_{0},\beta_{1}$ and the one loop anomalous dimension, and $C_{ij}$
are calculable constants depending on the higher order corrections
on $\beta(\alpha),\,\,\,\gamma(\alpha)$ and $O_{i}(\alpha)$. Note that
the summation within the bracket is not well defined; but, this point
will be irrelevant in the following discussion.

The corresponding Borel transform to (\ref{eq13}) is then given by
\begin{equation}
\tilde{A}(b) \approx \sum_{i}^{M}K_{i}
(1-\beta_{0}b)^{-\nu_{i}}\left[
1 + \sum_{j=1}^{\infty} C_{ij} (1-\beta_{0}b)^{j}\right]
\label{eq14}
\end{equation}
in the neighborhood of the singularity  at $b=1/\beta_{0}$,
and the corresponding large order behavior is given in the form
\begin{equation}
a_{n} = \sum_{i}^{M}\frac{ K_{i}}{\Gamma(\nu_{i})} 
 n! n^{\nu_i-1}\beta_{0}^{n} [ 1 +O(1/n)].
\end{equation}

Our aim is to express the prefactors $K_{i}$ in a calculable, convergent
series. Without losing generality, we may assume
$\nu_{i}>\nu_{j}$ for $i<j$. And also, for the moment we shall assume 
all $\nu_i >0$, and later  will make a comment on the case 
with a  negative $\nu_i$.
 Then it is straightforward to write
$K_{1}$, the prefactor of the most dominant term in the large order behavior,
in a convergent series. Since
\begin{equation}
K_{1}=\tilde{A}(b)
(1-\beta_{0}b)^{\nu_{1}}\left.\right|_{b=\frac{1}{\beta_{0}}},
\end{equation}
we can obtain a series expression for $K_{1}$ by expanding  the
function on the r.h.s. at $b=0$. The resulting series is then convergent
because the function is finite at the singularity and there is no other
singularity within the disk $|b| \leq -1/\beta_{0}$.

For the prefactors other than $K_{1}$ we  get a linear relation
among them.
Using (\ref{eq14}) it is easy to write $K_{i}$ as 
\begin{equation}
K_{i}=[ h_{i}(b) + \sum_{j}m_{ij}(b) K_{j}]
\left.\right|_{b=\frac{1}{\beta_{0}}}
\label{eq17}
\end{equation}
where
\begin{equation}
h_{i}(b)= \tilde{A}(b) (1-\beta_{0}b)^{\nu_{i}}
\end{equation}
and
\begin{eqnarray}
m_{ij}= \left\{ \begin{array}{l} 
 - (1-\beta_{0}b)^{\nu_{i}-\nu_{j}}
\left[1+\sum_{k=1}^{\left[ \nu_{j}-\nu_{i}\right]}
C_{jk}(1-\beta_{0}b)^{k}\right],
\,\,\,\,\,\mbox{for}\,\,\, i>j \\ 
0,\,\,\,\,\, \mbox{for}\,\,\, i\leq j \,, 
\end{array}\right.
\label{eq99}
\end{eqnarray}
with $\left[ \nu_{j}-\nu_{i}\right]$ being an integer satisfying
\begin{equation}
0 \leq \nu_{j}-\nu_{i} -\left[ \nu_{j}-\nu_{i}\right] <1.
\end{equation}

To solve  eq. (\ref{eq17}) we introduce  $r_i(b)$ defined by
\begin{equation}
r_i(b)=\sum_{j}[\delta_{ij} -m_{ij}(b)] K_j -h_i(b).
\label{eq100}
\end{equation}
Note that by definition $r_i(b)$ vanishes at the renormalon singularity.
From (\ref{eq100}) we obtain
\begin{equation}
K_i= \sum_{j}[1 -m^{(N)}]^{-1}_{ij}(h_{j}^{(N)} + r_{j}^{(N)})
\label{eq103}
\end{equation}
where $f^{(N)}$ for a function $f(b)$ denotes the $N$-th order Taylor expansion
evaluated at the singularity, i.e, 
\begin{equation}
f^{(N)}= \sum_{n=0}^{N} \frac{d^n f(0)}{db^n}\,\frac{\beta_{0}^{-n}}{n!}.
\end{equation}
Note that $h_{i}^{(N)}$ is calculable in terms of the perturbative
coefficients of $\tilde{A}(b)$.
From the definition (\ref{eq99}) we have
\begin{eqnarray}
m_{ij}^{(N)}= \left\{ \begin{array}{l} 
 - \frac{N^{\nu_{j}-\nu_{i}}}{\Gamma(\nu_j-\nu_i+1)}
\left[1+\sum_{k=1}^{\left[\nu_j-\nu_i\right]}\frac{
C_{jk}\Gamma(\nu_j-\nu_i+1)}{\Gamma(\nu_j-\nu_i-k+1)} N^{-k}\right],
\,\,\,\,\,\mbox{for}\,\,\, i>j \\ 
0,\,\,\,\,\, \mbox{for}\,\,\, i\leq j  
\end{array}\right.
\label{eq101}\end{eqnarray}
and from (\ref{eq14}), (\ref{eq100})
\begin{eqnarray}
r_i^{(N)}=&&\hspace{-0.15in}-\sum_{j=1}^{i-1} \sum_{k=[\nu_{j}-\nu_{i}]+1}^{\infty}
\frac{K_j C_{jk}}{\Gamma(\nu_{j}-\nu_{i}-k+1)} N^{\nu_j-\nu_i-k} \nonumber \\
&&\hspace{-0.15in}- \sum_{j=i+1}^{M} \frac{K_j}{\Gamma(\nu_j-\nu_i+1)}
N^{\nu_j-\nu_i}\left[1+
\sum_{k=1}^{\infty} \frac{ C_{jk}
\Gamma(\nu_j-\nu_i+1)}{\Gamma(\nu_j-\nu_i-k+1)}
N^{-k}\right]  \nonumber \\
&&\hspace{-0.15in}+\,\,  c_i N^{-\nu_i}\,( 1+ O(1/N))
\label{eq102}
\end{eqnarray}
where $c_{i}$ denotes a constant.
The  terms proportional to $c_i$ arise from
the regular part of the Borel transform  and are not
generally calculable. As expected $r_i^{(N)}$ vanishes in the large
$N$ limit.

Since $m_{ij}^{(N)}$ can be divergent under large $N$ limit,
$\sum_{j}(1-m^{(N)})^{-1}_{ij} r_j^{(N)}$ can be nonvanishing in large $N$ limit,
and thus we may write part of (\ref{eq103})
as
\begin{equation}
\sum_{j}[1 -m^{(N)}]^{-1}_{ij} r_{j}^{(N)} = \sum_{j}(\Delta_{1})_{ij}K_{j} + 
 r^1_i
\label{eq26}
\end{equation}
where we have separated those terms nonvanishing under large $N$ limit from
those vanishing, and put the former into
$\sum_{j}(\Delta_{1})_{ij}K_{j}$ and the
latter into  $ r^1_i $.
Note that the nonvanishing part is linear
in $K_i$. This is because the terms proportional to $c_i$ in (\ref{eq102})
give rise to a
contribution only of $O(N^{-\nu_{i}})$, thus affecting only 
$  r^1_i$,
which can be easily seen from
the fact that $(1-m^{(N)})^{-1}_{ij}$ is at most as divergent
as $m_{ij}^{(N)}$ in the large $N$ limit.
Substituting (\ref{eq26}) into (\ref{eq103}) we obtain
\begin{equation}
K_{i}= \sum_{j}[(1-\Delta_1)^{-1}(1 -m^{(N)})^{-1}]_{ij}h_j^{(N)} +
\sum_{j}(1-\Delta_1)^{-1}_{ij}  r^1_j.
\label{eq27}
\end{equation}
We can now repeat this step by
\be
\sum_{j}(1-\Delta_{m-1})^{-1}_{ij}r^{m-1}_j = \sum_{j}(\Delta_m)_{ij}K_{j} +
r^m_i
\ee
for a finite number of times (say $l$ times) until
$\sum_{j}(1-\Delta_l)^{-1}_{ij} r^l_j $ vanishes in the large $N$ limit,
to obtain
\begin{equation}
\vec{K} =\lim_{N\rightarrow\infty}\left[
(1-\Delta_l)^{-1}\cdots(1-\Delta_1)^{-1}
(1-m^{(N)})^{-1}\vec{h}^{(N)}\right].
\label{eq28}
\end{equation}
This is our main result.

Now as an example, let us consider a Borel transform whose  UV renormalon
singularity  is given by  (\ref{eq14}) with $M=2$. Then
from (\ref{eq101}) and (\ref{eq102})
\begin{equation}
m_{21}^{(N)}= -\frac{N^{\nu_1-\nu_2}}{\Gamma(\nu_1-\nu_{2}+1)} \,(1+O(1/N)), 
\hspace{.25in} \mbox{otherwise}\hspace{.25in} m_{ij}^{(N)}=0
\end{equation}
and
\begin{eqnarray}
r_1^{(N)}&=& -\frac{K_2}{\Gamma(\nu_2-\nu_1+1)} N^{\nu_2-\nu_1}\,(1+O(1/N)) +
c_1 N^{-\nu_1} \,(1+O(1/N)) \nonumber \\
r_2^{(N)}&=& -\frac{K_1 C_{11}}{\Gamma(\delta+1)}
N^{\delta}(1+O(1/N)) +c_2 N^{-\nu_2}\,(1+O(1/N))
\end{eqnarray}
where $\delta=\nu_1-\nu_2-[\nu_1-\nu_2]-1 < 0$.
Using the definition (\ref{eq26}) we find
\begin{equation}
\Delta_1= \left( \begin{array}{cc}
                 0 & 0 \\
		 0& \frac{1}{\Gamma(\nu_1-\nu_2+1)\Gamma(\nu_2-\nu_1+1)}
		 \end{array}
		 \right),
\end{equation}
and obtain
\begin{equation}
\left(\begin{array}{c}
      K_1 \\
      K_2 
      \end{array}
      \right) = \lim_{N\rightarrow\infty} \left[ 
(1-\Delta_1)^{-1}(1-m^{(N)})^{-1} 
\left(\begin{array}{c}
h_1^{(N)} \\
h_2^{(N)}
\end{array}\right)\right].
\end{equation}

Now having given the series expression for the normalization constants,
a comment is  in order. In deriving (\ref{eq28}) we have  assumed
that all $\nu_i >0$. As long as the normalization
constants associated with positive $\nu_i$ only  are concerned, (\ref{eq28})
can be used
without modification. However, if some of $\nu_i$ are negative, we
have to  consider $\tilde{A}^{(p)}(b)$   instead of $\tilde{A}(b)$
defined as 
\begin{eqnarray}
\tilde{A}^{(p)}(b)&=&\frac{d^{p}}{d b^p} \tilde{A}(b) \nonumber \\
               &\approx& \sum_{i}^{M}K'_{i}
(1-\beta_{0}b)^{-\tilde{\nu}_{i}}\left[
1 + \sum_{j=1}^{\infty} C'_{ij} (1-\beta_{0}b)^{j}\right] \hspace{0.5in}
\mbox{for} \hspace{0.25in} b \approx \frac{1}{\beta_0} 
\end{eqnarray}
such that $p$ satisfies $\tilde{\nu}_i=\nu_i +p >0$ for all $\nu_i$.
Clearly $K_i$ are related linearly with $K'_i$, and so calculable
once $K'_i$ are known.  Now  $K'_i$  from  $\tilde{A}^{(p)}(b)$
can be obtained in a series form  using the steps
taken in (\ref{eq17})-- (\ref{eq28}), and is given by  (\ref{eq28})
with $\vec{K} \rightarrow \vec{K'}, \nu_i \rightarrow\tilde{\nu}_i$
and  $\tilde{A}(b) \rightarrow
\tilde{A}^{(p)}(b)$.

In conclusion, we have shown that the normalization constants of the 
large order behavior caused by an UV renormalon can be expressed, as in the
case of an IR renormalon, in a calculable, convergent series, and thus can be
computed to an arbitrary precision using the ordinary weak coupling
expansion. Considering that the calculation of the normalization constants
is equivalent to summing all sort of the higher order renormalon diagrams,
it is surprising that they can be computed from the usual perturbation
expansion.

\vspace{.3in}
\noindent
{\bf Acknowledgements:}
This work was supported in part by the Korean Science and
Engineering Foundation (KOSEF).

 \end{document}